\documentclass[aps,preprint]{revtex4}

\usepackage{amssymb}
\usepackage{amsmath}
\usepackage{epsfig}
\usepackage{epstopdf}
\usepackage{bm}
\usepackage{graphicx,epsfig}
\usepackage{mathrsfs}
\usepackage{dcolumn}
\usepackage{color}
\usepackage{natbib}
\usepackage{CJK}
\def\be{\begin{equation}}
\def\ee{\end{equation}}
\def\bea{\begin{eqnarray}}
\def\eea{\end{eqnarray}}

\allowdisplaybreaks[2]

\begin{document}

\title{Creating locally interacting Hamiltonians in the synthetic frequency dimension for photons}
\author{Luqi Yuan$^{1,*}$, Avik Dutt$^{2}$, Mingpu Qin$^{3}$, Shanhui Fan$^{2,\dag}$, and Xianfeng Chen$^{1,4,\ddag}$}
\affiliation{$^1$State Key Laboratory of Advanced Optical Communication Systems and Networks, School of Physics and Astronomy, Shanghai Jiao Tong University, Shanghai 200240, China \\
$^2$Department of Electrical Engineering, and Ginzton Laboratory,
Stanford University, Stanford, CA 94305, USA\\
$^3$School of Physics and Astronomy, Shanghai Jiao Tong University, Shanghai 200240, China \\
$^4$Collaborative Innovation Center of Light Manipulations and
Applications, Shandong Normal University, Jinan 250358£¬China\\
$^*$Corresponding author: yuanluqi@sjtu.edu.cn\\
$^\dag$Corresponding author: shanhui@stanford.edu\\
$^\ddag$Corresponding author: xfchen@sjtu.edu.cn}

\date{\today }

\begin{abstract}
The recent emerging field of synthetic dimension in photonics
offers a variety of opportunities for manipulating different
internal degrees of freedom of photons such as the spectrum of
light. While nonlinear optical effects can be incorporated into
these photonic systems with synthetic dimensions, these nonlinear
effects typically result in long-range interactions along the
frequency axis. Thus it has been difficult to use of the synthetic
dimension concept to study a large class of Hamiltonians that
involves local interactions. Here we show that a Hamiltonian that
is locally interacting along the synthetic dimension can be
achieved in a dynamically-modulated ring resonator incorporating
$\chi^{(3)}$ nonlinearity, provided that the group velocity
dispersion of the waveguide forming the ring is specifically
designed. As a demonstration we numerically implement a
Bose-Hubbard model, and explore photon blockade effect, in the
synthetic frequency space. Our work
 opens new possiblities
for studying fundamental many-body physics in the synthetic space
in photonics, with potential applications in optical quantum
communication and quantum computation.
\end{abstract}

\maketitle

The concept of synthetic dimension enables one to explore higher
dimensional physics in lower dimensional systems
\cite{boada12,jukic13,celi14,mancini15,stuhl15,gadway15,livi16,kolkowitz17,price17,martin17,baum18}.
In photonics \cite{opticareview,NRPreview}, synthetic dimension
can be achieved by exploiting various degrees of freedom of light
such as its frequency \cite{yuanol,ozawa16}, spatial mode
\cite{lustig19}, or orbital angular momentum \cite{luo15}, and by
specifically controlling the coupling between these degrees of
freedom \cite{schwartz13,yuanhaldane,lustigcleo19}.  While
initially explored mostly theoretically
\cite{yuanoptica,linnc,zhang17,luo17,zhou17,luo18,yuanprl19}, very
recently there have been a number of important experimental
developments, including the first experimental demonstration of
topological insulators with synthetic dimensions \cite{lustig19},
the measurement of band structure along the frequency axis of
light \cite{aviknc}, and the control of light spectrum in
synthetic space \cite{bell17,qin18}.

Most of the existing theoretical and experimental works on
synthetic dimension in photonics concerns linear processes without
photon-photon interactions. Certainly, it would be of interest to
consider nonlinear systems where there is photon-photon
interaction. Moreover, since one of the major objectives by going
into synthetic dimension is to create a platform to study specific
interacting Hamiltonians that are of physical significance, it
would be important to develop a strategy to synthesize these
interacting Hamiltonians. A very large class of interesting
interacting Hamiltonians have \textit{local} interactions
\cite{bloch08,chang14,roy17,abanin19}. Attempts to achieve such
locally interacting Hamiltonians in synthetic space and have been
made in photonics in the geometric angular coordinate
\cite{ozawa17} and also in ultracold atoms with spin being the
internal degree of freedom \cite{barbiero19}. On the other hand,
as is known and we will briefly reiterate below, the standard form
of nonlinear optics typically leads to a form of interaction that
is \textit{nonlocal} in the synthetic space
\cite{boydbook,strekalov16}. How to achieve a Hamiltonian with
local interactions in the synthetic space for photons, thus
therefore represents an important open theoretical question.

\begin{figure}[!h]
\centering
\includegraphics[width=1\linewidth]{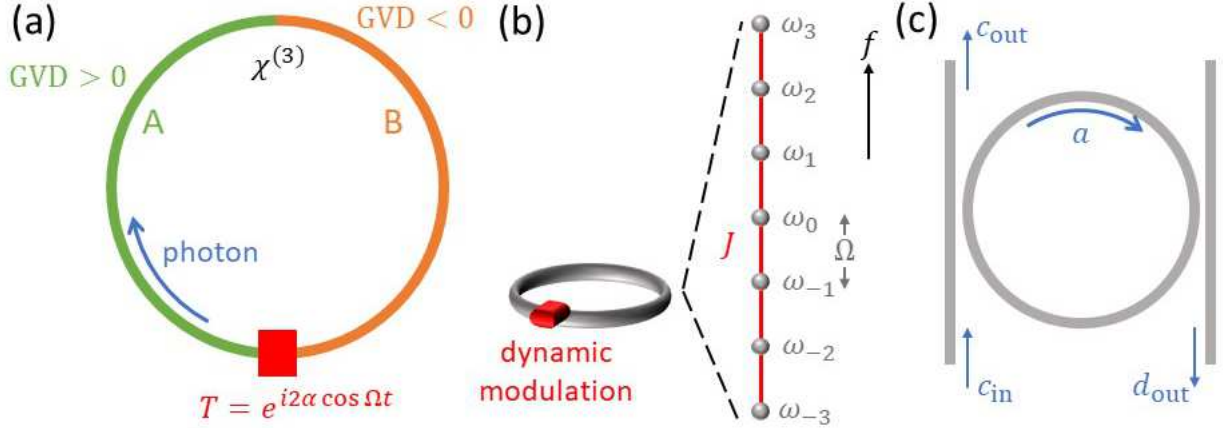}
\caption{(a) A ring resonator, composed by two types of
single-mode waveguides A and B, undergoing the dynamic modulation.
(b) A ring under the dynamic modulation supports a synthetic
lattice along the frequency dimension. (c) The ring is coupled
with the through-port and drop-port waveguides. \label{Fig:ring}}
\end{figure}

In this Letter, we consider a  synthetic frequency dimension of
light, formed in a ring resonator incorporating a modulator. To
create photon-photon interaction we consider $\chi^{(3)}$
processes in the waveguide forming the ring.  We show that a local
photon-photon interaction in the frequency dimension can be
achieved with a careful design of the group velocity dispersion of
the waveguide [Fig. \ref{Fig:ring}(a)]. As a demonstration we show
that this system can be used to demonstrate a Bose-Hubbard model,
and achieve photon-blockade effects along the synthetic axis. This
work here significantly broadens the range of physics phenomena
that can be studied in photonic synthetic space, and may lead to
new opportunities in quantum simulations and in the manipulation
of light.

We start with a brief review of the ring resonator under the
dynamic modulation as shown in Fig. \ref{Fig:ring}(b), which
naturally leads to a synthetic dimension of light along the
frequency axis. We consider a ring composed by a single-mode
waveguide with an effective refractive index $n_{\mathrm{eff}}$,
and a length $l$. For simplicity, assuming  zero group velocity
dispersion in the waveguide, the ring resonator supports resonant
modes at the resonant frequencies $\omega_m = \omega_0 + m \Omega$
where $\Omega \equiv 2\pi c/ n_{\mathrm{eff}}l \ll \omega_0$ is
the free-spectral-range (FSR) for the ring. We place a phase
modulator inside the ring and choose the modulation frequency to
be equal to the free spectral range. The transmission of light
passing through the modulator can be described by a time-dependent
transmission coefficient
\begin{equation}
T = e^{i 2 \alpha \cos(\Omega t)}, \label{Eq2:transmission}
\end{equation}
where $\alpha$ is the modulation amplitude. We assume that the
field experiences small changes for each round-trip. Then after
one round-trip, the change of the field amplitude $E_m$ for the
$m$-th resonant mode can be written as
\begin{equation}
\frac{2\pi}{\Omega}\frac{\partial E_m}{\partial T} \equiv \Delta
E_m = i \alpha (E_{m+1} + E_{m-1}), \label{Eq2:change}
\end{equation}
where $T$ is the slow time variable \cite{haus00}. The dynamics of
Eq. (\ref{Eq2:change}) is described by an effective Hamiltonian:
\begin{equation}
H_0 = - \hbar J \sum_m \left( a_m^\dagger a_{m+1} + h.c.\right),
\label{Eq2:hamiltonian}
\end{equation}
where $J = \alpha \Omega/2\pi$ is the modulation strength and
$a_m$ ($a_m^\dagger$) is the annihilation (creation) operator for
the $m$-th resonant mode. The Hamiltonian (\ref{Eq2:hamiltonian})
describes a one-dimensional tight-binding model along the
synthetic frequency axis [see Fig. \ref{Fig:ring}(b)]. With this
approach, a wide variety of non-interacting Hamiltonian with
different connectivity and topological properties can be created
\cite{opticareview}. However, there have been far less works on
creating important \textit{interacting} Hamiltonians.

\begin{figure}[!h]
\centering
\includegraphics[width=\linewidth]{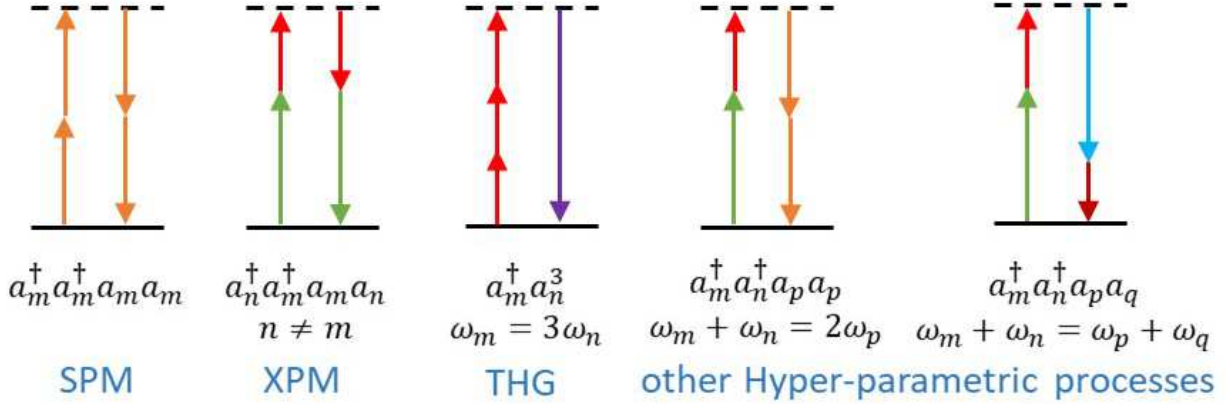}
\caption{Four-wave-mixing processes in a ring with a third-order
nonlinear susceptibility including the hyper-parametric processes
(SPM, XPM and others) and the THG process. \label{Fig:fwm}}
\end{figure}

As an  illustration of the general difficulty as well as our
approach of creating locally interacting Hamiltonian in the
frequency synthetic dimension, we aim to synthesize the effective
Bose-Hubbard Hamiltonian in the synthetic frequency dimension
\begin{equation}
H_1 = - \hbar J \sum_m \left( a_m^\dagger a_{m+1} + h.c.\right)
 -  \frac{\hbar g}{2} \sum_m a_m^\dagger a_m^\dagger a_m a_m , \label{Eq2:hamtarget}
\end{equation}
where $g$ is the local nonlinear strength. Although nonlinear
optical phenomena in many photonic materials have been extensively
studied, creating the Hamiltonian of Eq. (\ref{Eq2:hamtarget}) in
the frequency synthetic space is in fact non-trivial
\cite{boydbook}. The introduction of the nonlinearity typically
leads to long-range interactions over all synthetic lattice sites.
For example, the dynamically-modulated ring as shown in Fig.
\ref{Fig:ring}(b), with a third-order nonlinear susceptibility
$\chi^{(3)}$ is described by an  interacting Hamiltonian
\cite{strekalov16}:
\begin{equation}
H_2 = - \hbar J \sum_m \left( a_m^\dagger a_{m+1} + h.c.\right) -
\frac{\hbar g}{2} \sum_{m,n,p,q} a_m^\dagger a_n^\dagger a_p a_q -
\frac{\hbar g}{3} \sum_{m,n} \left( a_m^\dagger a_n^3 + h.c.
\right), \label{Eq1:nonlinear}
\end{equation}
as can be derived using the rotating wave approximation. In Eq.
(\ref{Eq1:nonlinear}), the second and third terms describe two
four-wave-mixing effects. The second term describes the
hyper-parametric oscillation process involving four modes with
frequencies with $\omega_m+\omega_n = \omega_p+\omega_q$. The
third term describes the third harmonic generation (THG) process
with $\omega_m = 3\omega_n$. Comparing Eq. (\ref{Eq2:hamtarget})
with Eq. (\ref{Eq1:nonlinear}), we see that the interacting term
in Eq. (\ref{Eq2:hamtarget}) corresponds to the self-phase
modulation (SPM) process. In Eq. (\ref{Eq1:nonlinear}), however,
in addition to the SPM process,  other terms which describe  the
cross phase modulation (XPM) process, other hyper-parametric
processes, and the THG process result in   long-range interactions
between different resonant modes in the ring (see Fig.
\ref{Fig:fwm}). While here for illustration purposes we consider a
specific type of nonlinearity, the observation is in fact rather
general: standard nonlinearity does not lead to local interactions
in the synthetic frequency dimension.

In order to synthesize the Bose-Hubbard model where the
interaction is local  along the frequency dimension, we propose a
ring consisting of two sections of waveguides. The two sections
have the same $\chi^{(3)}$ nonlinear susceptibility. However, they
have opposite group velocity dispersion (GVD), as illustrated in
Fig. \ref{Fig:ring}(a). In the linear regime of light propagation,
for each complete round-trip as light goes around the ring, the
dispersion effect in the two waveguide sections cancels. Hence
this ring also features equally spaced resonances along the
frequency axis, same as a ring with zero GVD, and a
one-dimensional synthetic frequency dimension can be created in
this ring by applying the dynamic modulation. On the other hand,
in each waveguide section, due to the the group velocity
dispersion, the only phase-matched processes are the SPM and XPM
processes. Moreover, as we will show below, in any Hilbert space
with a fixed photon number the Hamiltonian for the XPM process in
fact reduces to that for the SPM process. We therefore show that
our design supports an effective Bose-Hubbard Hamiltonian
(\ref{Eq2:hamtarget}) along the frequency axis.

In the following, we explain our design in details. We first
consider light propagating inside a single-mode waveguide. The
evolution of the field amplitude $A(z,\omega)$ with the frequency
$\omega$ at the position $z$ in the waveguide can be described as
\cite{hausbook}:
\begin{equation}
\frac{\partial A(z,\omega)}{\partial z} = -i \beta(\omega)
A(z,\omega), \label{Eq1:propagation1}
\end{equation}
where $\beta(\omega)$ denotes the propagation wavevector.
$\beta(\omega)$ can be expanded around the reference frequency
$\omega_0$, i.e.,
\begin{equation}
\beta (\omega) - \beta (\omega_0) =
\left.\frac{d\beta}{d\omega}\right|_{\omega_0}
\left(\omega-\omega_0\right) +
\frac{1}{2}\left.\frac{d^2\beta}{d\omega^2}\right|_{\omega_0}
\left(\omega-\omega_0\right)^2 + \ldots . \label{Eq1:betaexpand}
\end{equation}
Here we set $\beta_0 = \beta (\omega_0)$, $v_g =
\left(d\beta/d\omega\right)^{-1}$ is the group velocity, and
$\beta_2 = d^2\beta/d\omega^2$ is the GVD around $\omega_0$. By
neglecting the higher-order terms in Eq. (\ref{Eq1:betaexpand}),
one can write the field amplitude of light after it propagates
through a waveguide with the length $L$
\begin{equation}
A(z=L,\omega) = A(z=0,\omega) e^{-i \beta_0 L -i
\left(\omega-\omega_0\right) L/v_g - i \beta_2
\left(\omega-\omega_0\right)^2 L/2} . \label{Eq1:field}
\end{equation}
Note that the GVD induces a frequency-dependent phase delay.

\begin{figure}[!h]
\centering
\includegraphics[width=\linewidth]{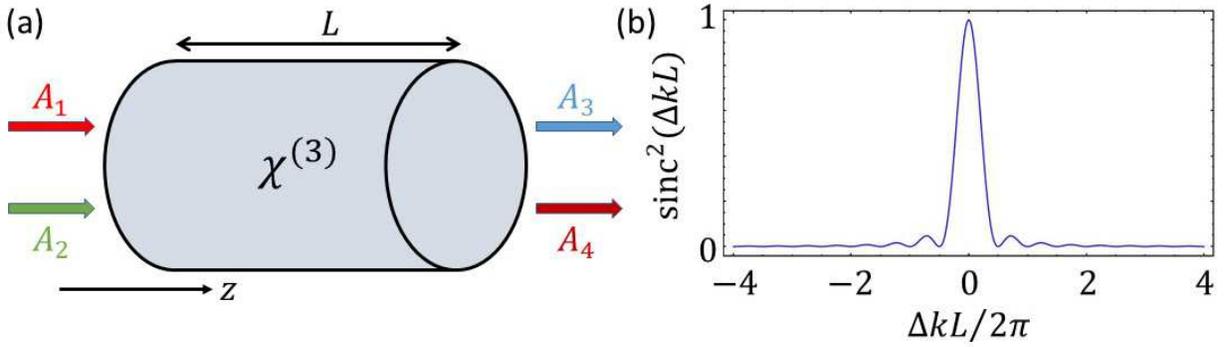}
\caption{(a) A general four-wave-mixing process in a waveguide
with the third-order susceptibility $\chi^{(3)}$. (b) The plot of
$\Delta I \propto \mathrm{sinc}^2(\Delta k L)$ showing the effect
of the phase mismatching  . \label{Fig:fwmsinc}}
\end{figure}

Next, we explore the light propagates through the waveguide
section having the positive GVD ($\beta_2 > 0$) and  $\chi^{(3)}$
as shown in Fig. \ref{Fig:ring}(a). For simplicity, we consider
two incident waves with the frequencies $\omega_1$ and $\omega_2$
respectively, in the vicinity of the reference frequency
$\omega_0$. We look for the solution for a general
four-wave-mixing process where output fields are generated  at
frequencies $\omega_3$ and $\omega_4$, satisfying the energy
conservation relation $\omega_3 + \omega_4 = \omega_1 + \omega_2$
[see Fig. \ref{Fig:fwmsinc}(a)]. We also assume the narrowband
limit, i.e., $|\omega_0 - \omega_m|\ll \omega_0, m=1,2,3,4$. The
resulting coupled-mode equations under the slowly varying
amplitude approximation are \cite{boydbook}
\begin{equation}
\frac{\partial \tilde A_3}{\partial z} =
i\frac{\omega_0^2}{2\beta_0 c^2} \chi^{(3)} \tilde A_1 \tilde A_2
\tilde A_4^* e^{-i\Delta k z}, \label{Eq1:propagation23}
\end{equation}
\begin{equation}
\frac{\partial \tilde A_4}{\partial z} =
i\frac{\omega_0^2}{2\beta_0 c^2} \chi^{(3)} \tilde A_1 \tilde A_2
\tilde A_3^* e^{-i\Delta k z}, \label{Eq1:propagation24}
\end{equation}
where $\tilde A_m = A_m e^{i \beta(\omega_m) z}$ and the momentum
mismatch is
\begin{equation}
\Delta k = \frac{\beta_2}{2} \left[
\left(\omega_3-\omega_0\right)^2 +
\left(\omega_4-\omega_0\right)^2 -
\left(\omega_1-\omega_0\right)^2 -
\left(\omega_2-\omega_0\right)^2 \right] . \label{Eq1:mismatch}
\end{equation}
At frequencies around  $\omega_0$, the phase matching condition
($\Delta k =0$) is satisfied if $\omega_1 = \omega_2 = \omega_3 =
\omega_4$, which corresponds to  the SPM  process, or $\omega_1 =
\omega_{3(4)}$ and $\omega_2 = \omega_{4(3)} \neq \omega_1$, which
corresponds to  the XPM  process. For other FWM processes with the
phase mismatching condition $\Delta k \neq 0$, the efficiency for
frequency conversion  is proportional to the phase mismatch
factor, i.e., $\mathrm{sinc}^2(\Delta k L)$, as plotted in  Fig.
\ref{Fig:fwmsinc}(b). For $|\Delta k L| \geq \pi$, the conversion
efficiency is significantly suppressed. In particular, for
$|\Delta k L|=n\pi$ with $n$ being the positive integer,
$\mathrm{sinc}^2(\Delta k L)=0$, meaning that FWM processes other
than SPM and XPM vanish completely. Similar argument is also valid
for the THG process, where the conversion efficiency of THG is
also strongly reduced due to the phase mismatching.

Having treated a light propagation effect in a single waveguide
section, we now discuss the ring in Fig. \ref{Fig:ring}(a). The
waveguide A with the length $L=l/2$ has a positive GVD
($\beta_{2,\mathrm{A}} >0$). The waveguide B with the length $L$
has a negative GVD ($\beta_{2,\mathrm{B}}<0$), we assume that
$\beta_{2,\mathrm{B}} = - \beta_{2,\mathrm{A}}$. The quadratic
term of the phase delay in Eq. (\ref{Eq1:field}) due to the
dispersion effect is therefore compensated once the light
propagates through the entire ring. Hence this ring supports
equally-spaced resonant modes at the resonant frequencies
$\omega_m$ with the FSR $\Omega$. As light goes through each round
trip, its amplitude experiences a small change due to both the
nonlinearity and the dynamic modulation. After a round-trip, the
change of the field amplitude $\tilde A_m$ for the $m$-th mode at
frequency $\omega_m$ can be written from Eqs. (\ref{Eq2:change}),
(\ref{Eq1:propagation23}), and (\ref{Eq1:propagation24})
\begin{equation}
\frac{2\pi}{\Omega}\frac{\partial\tilde A_m}{\partial T}  = i
\alpha (\tilde A_{m+1} + \tilde A_{m-1}) + i
\frac{\omega_0^2}{2\beta_0 c^2} \chi^{(3)} |\tilde A_m|^2 \tilde
A_m + i \sum_{n\neq m} \frac{2\omega_0^2}{\beta_0 c^2} \chi^{(3)}
|\tilde A_n|^2 \tilde A_m, \label{Eq1:change}
\end{equation}
The corresponding effective Hamiltonian from Eq.
(\ref{Eq1:change}) reads:
\begin{equation}
H_3 = - \hbar J \sum_m \left( a_m^\dagger a_{m+1} + h.c.\right)
 -  \frac{\hbar g}{2} \sum_m a_m^\dagger a_m^\dagger a_m a_m - \hbar g \sum_{m,n\neq m} a_n^\dagger a_m^\dagger a_m a_n. \label{Eq1:hamiltonian}
\end{equation}
Here $g = 3\hbar \omega_0^2 \chi^{(3)}/V n_0^4 \epsilon_0$
describes the strength of the nonlinearity
\cite{matsko05,chembo10}, where $V$ is the effective mode volume,
$n_0 = n(\omega_0)$, and $\epsilon_0$ is the permittivity of free
space.

In a Hilbert space where the total number of photons $N$ is
conserved in the the system,  Eq. (\ref{Eq1:hamiltonian}) can be
further simplified by noting that
\begin{equation}
\sum_{m,n\neq m} a_n^\dagger a_m^\dagger a_m a_n = \sum_m
a_m^\dagger a_m \left(\sum_{n\neq m} a_n^\dagger a_n\right) =
\sum_m a_m^\dagger a_m \left(N - a_m^\dagger a_m \right) = N^2 - N
- \sum_m a_m^\dagger a_m^\dagger a_m a_m. \label{Eq1:derivation}
\end{equation}
The resulting Hamiltonian reads
\begin{equation}
H_4 = - \hbar J \sum_m \left( a_m^\dagger a_{m+1} + h.c.\right)
 +  \frac{\hbar g}{2} \sum_m a_m^\dagger a_m^\dagger a_m a_m - \hbar g (N^2-N). \label{Eq1:hamiltonian2}
\end{equation}
One notices that the third term $\hbar g (N^2-N)$ is just a
constant shift in energy. The change of sign in the second term
does not affect the physical observables for bosonic systems.
Hence, the Hamiltonian (\ref{Eq1:hamiltonian2}) [or the
Hamiltonian (\ref{Eq1:hamiltonian})] describes the same physics as
our desired Hamiltonian in Eq. (\ref{Eq2:hamtarget}). We have
therefore shown that a Bose-Hubbard model with local interaction
in the synthetic frequency dimension can be achieved by the ring
shown in  Fig. \ref{Fig:ring}(a).

To demonstrate some of the physics effects along the synthetic
frequency dimension in the Hamiltonian (\ref{Eq1:hamiltonian}), we
perform numerical simulations with photon number $N=2$. External
waveguides are coupled with the ring, where the photons input into
the ring from the through-port and the detection can be made at
the drop-port waveguide [see Fig. \ref{Fig:ring}(c)]. Hence, our
system becomes open, and the photon transport property is
described by the input-output formalism in the Heisenberg picture
\cite{gardiner85,fan10}:
\begin{equation}
\frac{d a_m(t)}{dt} = \frac{i}{\hbar}[H_3, a_m] - \gamma a_m (t) +
i \sqrt{\gamma} c_{\mathrm{in},m} (t), \label{Eq1:inputoutput1}
\end{equation}
\begin{equation}
c_{\mathrm{out},m} (t) = c_{\mathrm{in},m} (t) - i\sqrt{\gamma}
a_m(t), \label{Eq1:inputoutput2}
\end{equation}
\begin{equation}
d_{\mathrm{out},m} (t) =  - i\sqrt{\gamma} a_m(t),
\label{Eq1:inputoutput3}
\end{equation}
where $\gamma$ is the waveguide-cavity coupling strength.
$c_{\mathrm{in},m}$ ($c_{\mathrm{out},m}$) and
$d_{\mathrm{out},m}$ are the input (out) annihilation operators
for photons at the frequency $\omega_m$ in the through-port and
drop-port waveguides, respectively.

As the input state, we consider a strongly-correlated photon pair
\cite{yuananyon}:
\begin{equation}
| \phi (p,q) \rangle =  \iint dt_1 dt_2
f\left(\frac{t_1+t_2}{2}\right) h\left(t_1-t_2\right)
c_{\mathrm{in},q}^\dagger (t_2) c_{\mathrm{in},p}^\dagger (t_1)
|0\rangle, \label{Eq1:inputphoton}
\end{equation}
which satisfies the normalization condition $\langle \phi | \phi
\rangle=1$. We further assume  that $h(t)$ has an extremely short
temporal width. Hence the input state (\ref{Eq1:inputphoton})
describes the scenario where two photons at the frequencies
$\omega_p + \Delta \omega$ and $\omega_q + \Delta \omega$ are
simultaneously injected into the through-port waveguide. We choose
$f(t) = e^{ - (t - t_D)^2/ \Delta t^2 } e^{-i 2\Delta \omega t}$,
where $t_D$ describes the timing of the incident photons, $\Delta
t$ is the pulse temporal width, and $\Delta \omega$ is the small
frequency detuning. To excite the individual sites in the
synthetic lattice along the frequency dimension, the condition
$1/\Delta t \ll \Omega$ is required.

To measure the quantum statistics of the photons inside the ring,
we define the two-photon correlation function $G^{(2)}_{m,n}
(t,t') =\langle \phi | d_{\mathrm{out},m}^\dagger (t)
d_{\mathrm{out},n}^\dagger (t') d_{\mathrm{out},n} (t')
d_{\mathrm{out},m} (t) | \phi \rangle$ \cite{scullybook}. We then
compute the two-photon correlation probability that two output
photons coincide in time
\begin{equation}
P_{m,n} =  \int dt  G^{(2)}_{m,n} (t,t). \label{Eq1:detection}
\end{equation}

\begin{figure}[!h]
\centering
\includegraphics[height=0.3\linewidth]{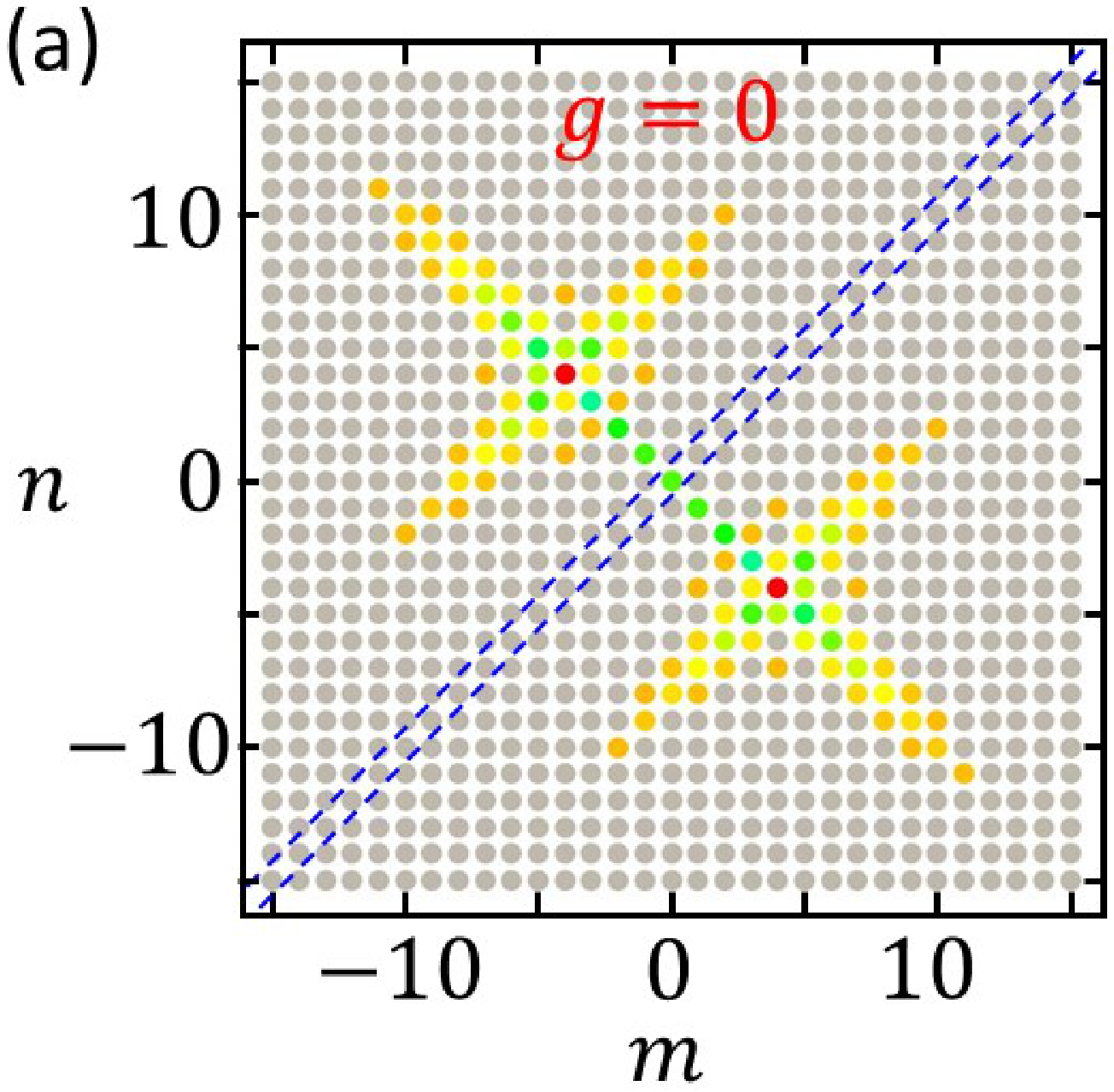}
\includegraphics[height=0.3\linewidth]{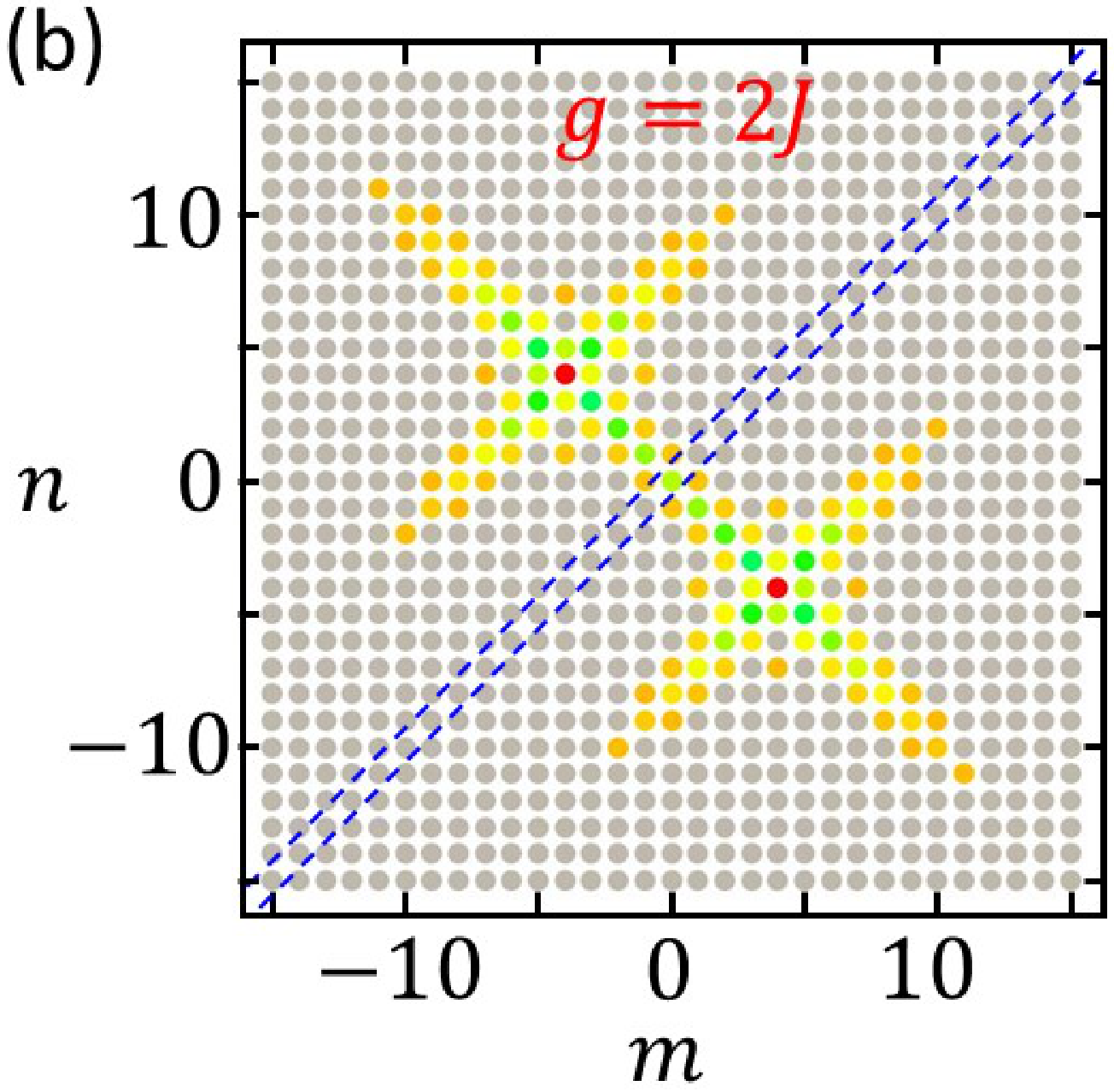}
\includegraphics[height=0.3\linewidth]{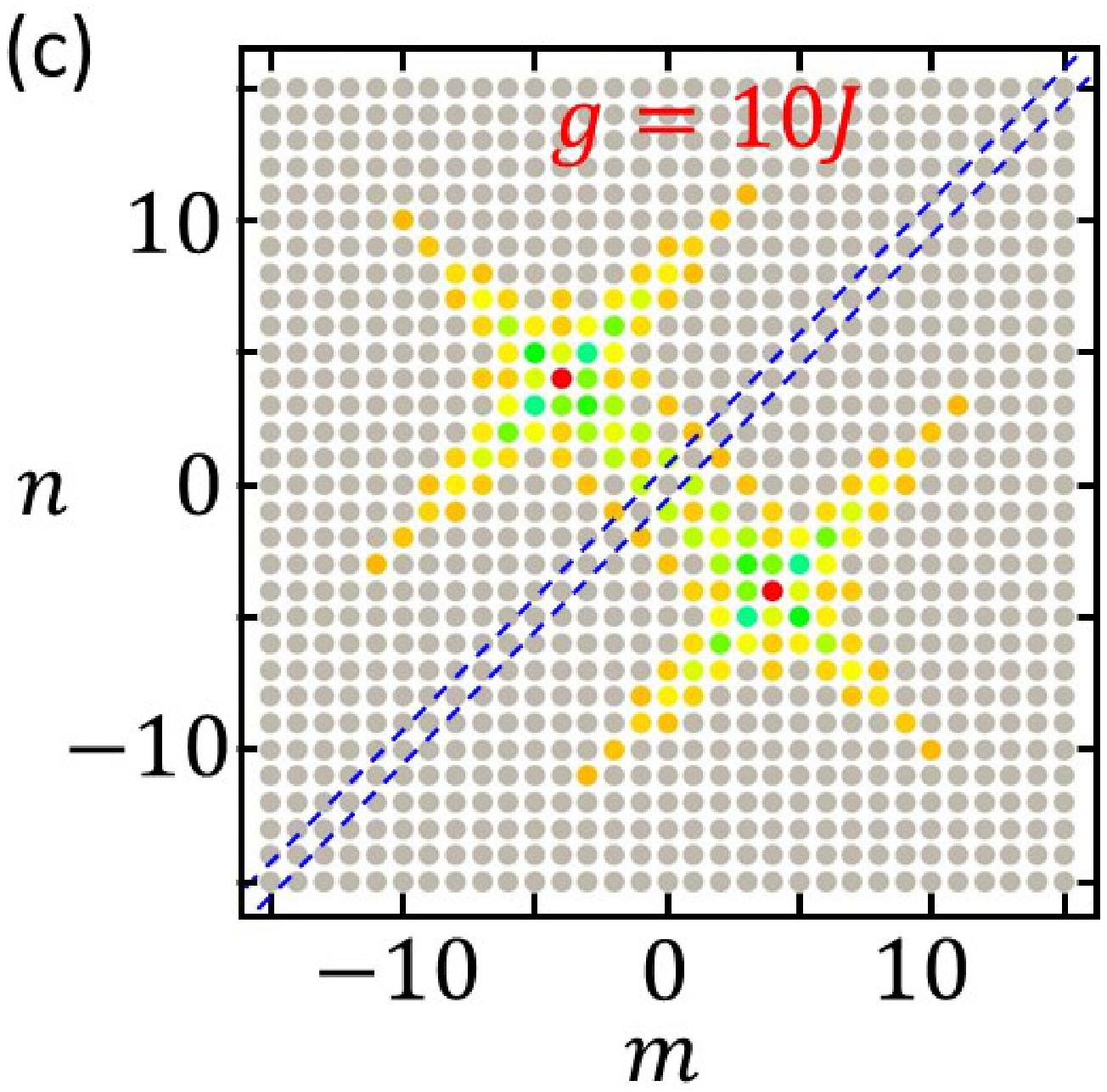}
\includegraphics[height=0.3\linewidth]{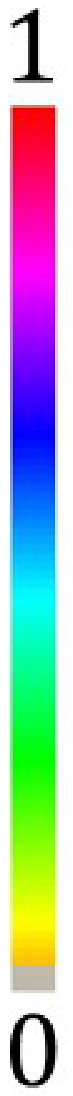}
\includegraphics[height=0.3\linewidth]{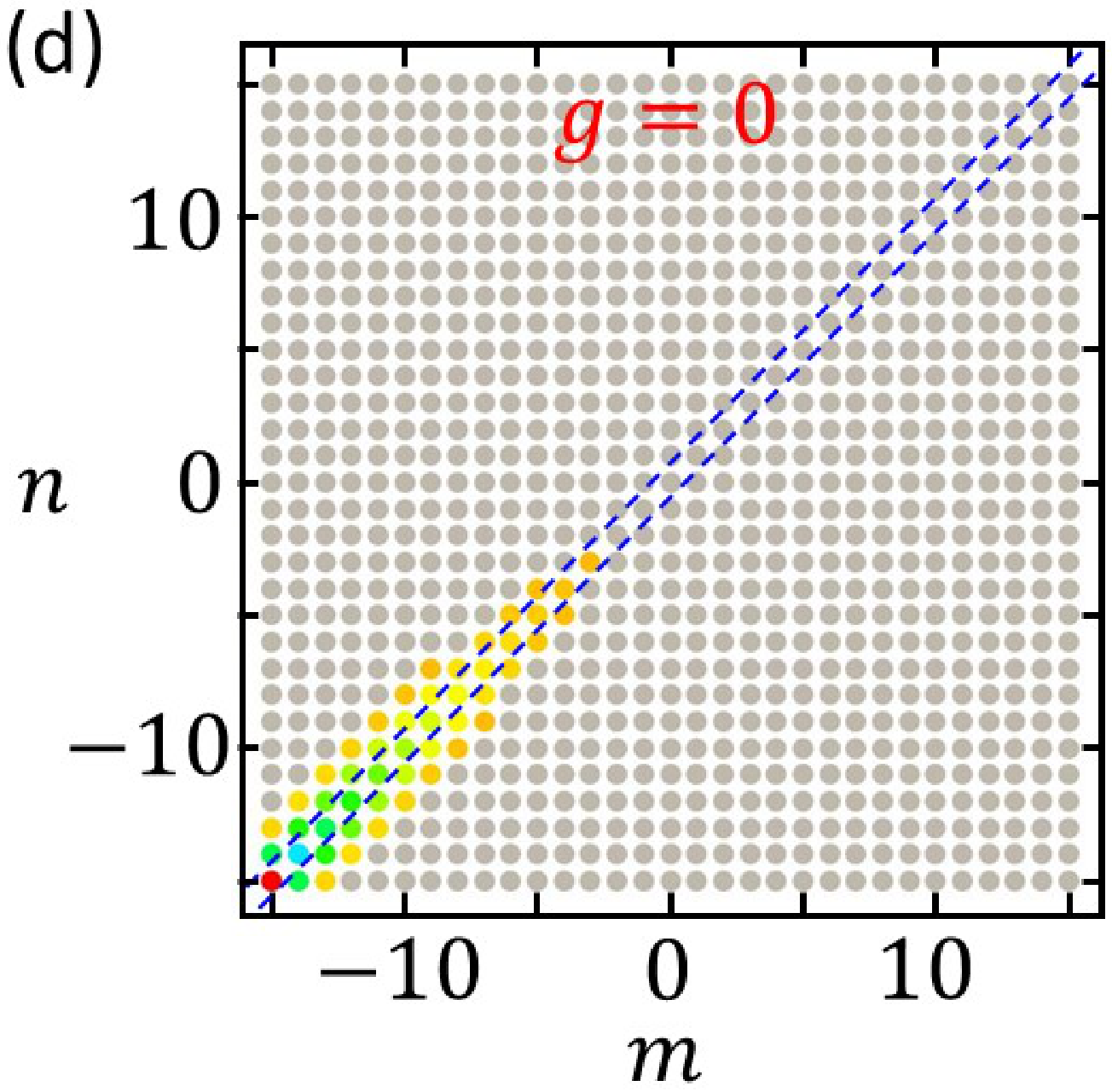}
\includegraphics[height=0.3\linewidth]{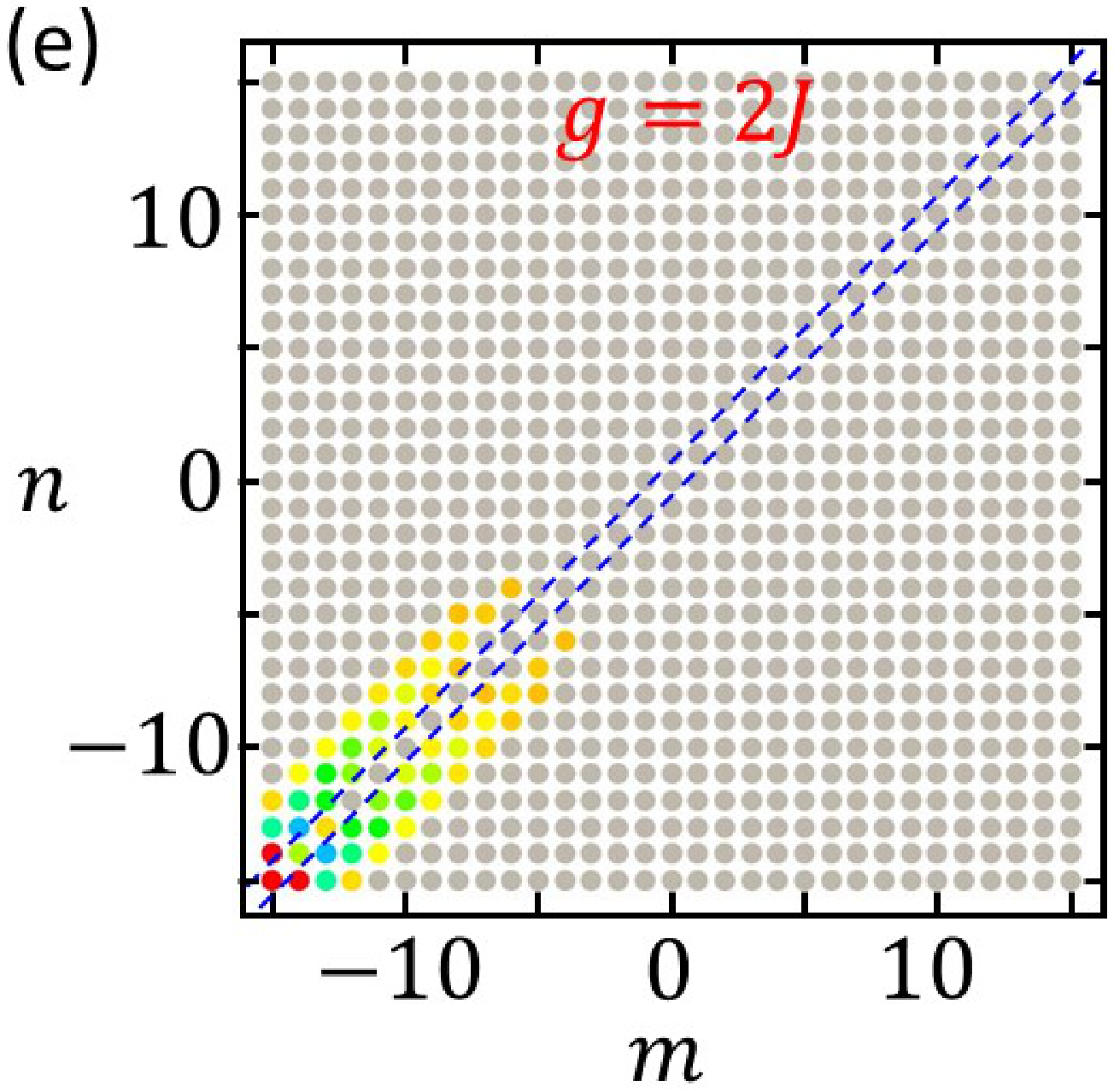}
\includegraphics[height=0.3\linewidth]{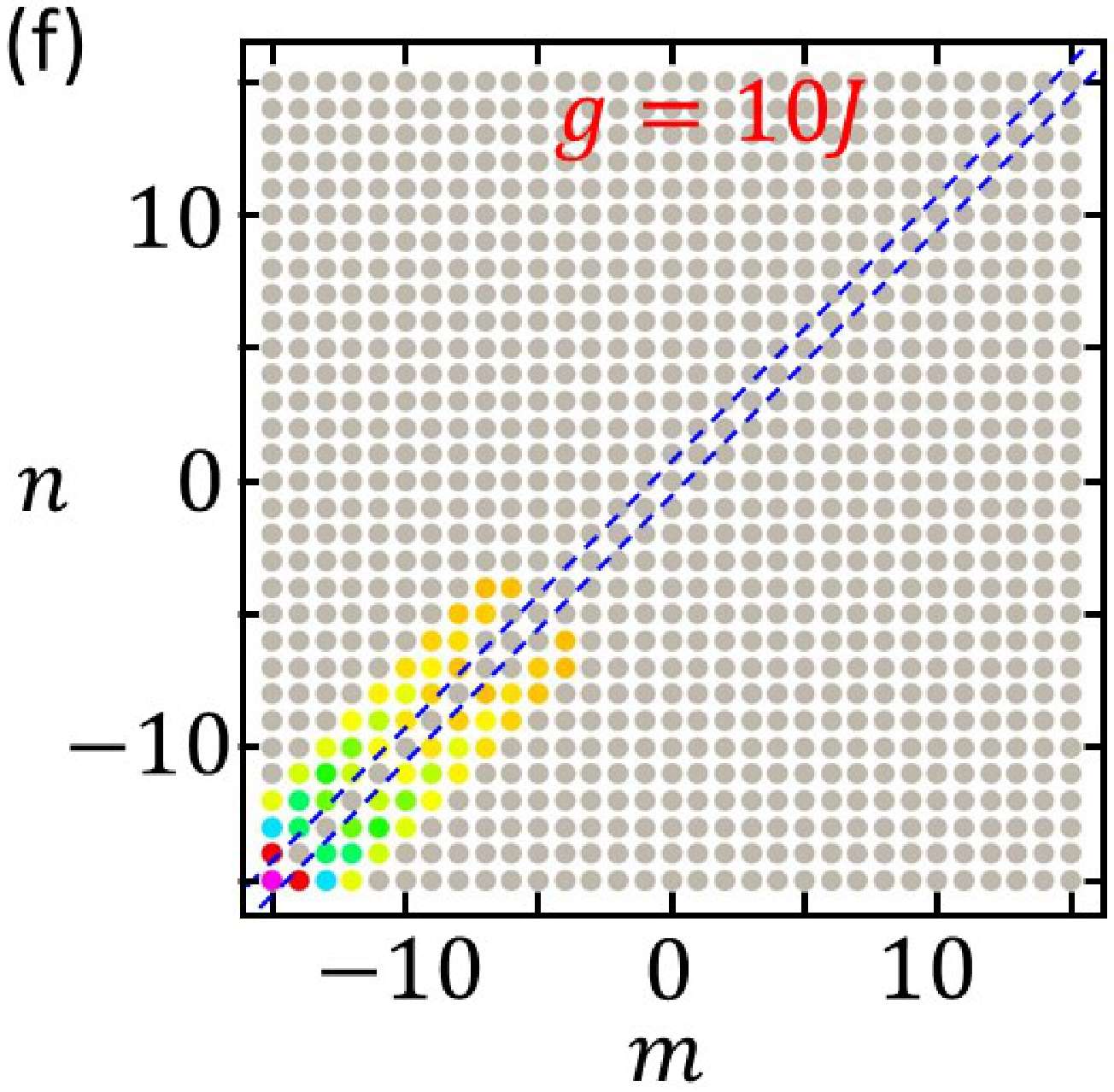}
\includegraphics[height=0.3\linewidth]{legend.eps}
\caption{Normalized distributions of the two-photon correlation
probability $P_{m,n}$. (a)-(c) The input photon pair is $| \phi
(-4,4) \rangle$ with $g=0$, $g=2J$, and $g=10J$, respectively.
(a)-(c) The input photon pair is $| \phi (-15,-15) \rangle$ with
$g=0$, $g=2J$, and $g=10J$, respectively. Positions inside two
dashed lines corresponds to $P_{m,m}$. \label{Fig:sim}}
\end{figure}

The simulation procedure follows the standard formalism
\cite{yuananyon} but in the synthetic frequency dimension. We
consider the synthetic lattice in Fig. \ref{Fig:ring}(b) involving
31 resonant modes ($m=-15,\ldots,15$), and set $\gamma = 0.2J$,
$t_D = 20 J^{-1}$, and $\Delta t = 6.4 J^{-1}$. We also choose
$\Delta \omega =  g$ to compensate for the overall  frequency
shift in the effective Bose-Hubbard model. A photon pair $| \phi
(-4,4) \rangle$ is injected into the waveguide. We consider the
cases with  $g=0$, $g=2J$, and $g=10J$ in the simulations.
Normalized distributions of the two-photon correlation probability
$P_{m,n}$ for each choice of $g$ are plotted in Figs.
\ref{Fig:sim}(a)-(c). There is no photon interaction in the $g =
0$ case. Notice the existence of non-zero probability $P_{m,m}$
where the output photons have the same frequency. As the
interaction increases the output photon statistics changes.  For a
large $g=10J$, the correlation probability $P_{m,m}=0$ for all
$m$. High nonlinearity thus introduces a large blockade effect at
each resonant mode along the synthetic frequency dimension.

We next use the photon pair $| \phi (-15,-15) \rangle$ as the
input and assume that there is a sharp boundary at $\omega_{-16}$
in the synthetic frequency dimension. Such an artificial sharp
boundary can be created by adding another ring to strongly couple
only to the $-16$th resonant mode  but no other modes
\cite{yuanprl19}. In this input state the two photons have the
same carrier frequencies. Figs. \ref{Fig:sim}(d)-(f) show the
simulated normalized distributions of the two-photon correlation
probability $P_{m,n}$ for different $g$. At $g = 0$, there is a
strong probability for the two photons to have the same
frequencies. In contrast, with a large nonlinearity $g = 10J$,
away from the input frequency at $\omega_{-15}$, there is very
little probability that the output photons exhibit the same
frequencies. Again, we see strong photon-blockade effect in the
synthetic frequency dimension.

Simulations in Fig. \ref{Fig:sim} show the photon-photon
interaction along the synthetic frequency dimension, which is an
analog to the on-site nonlinear interaction in the spatial
dimension \cite{birnbaum05,gorshkov11}. Moreover, we note that
simulation results in Fig. \ref{Fig:sim} are the same as results
performed by simulating Eqs.
(\ref{Eq1:inputoutput1})-(\ref{Eq1:inputoutput3}) with $H_3$
replaced by  $H_1$ in Eq. (\ref{Eq2:hamtarget}), indicating that
the Hamiltonian $H_3$ in Eq. (\ref{Eq1:hamiltonian}) indeed
provides an implementation of the Bose-Hubbard model  in the
synthetic frequency dimension.

Our proposed system requires waveguides with strong nonlinearity
and large group velocity dispersion, which is experimentally
challenging but may be achievable with the current
start-of-the-art technology. For a ring with the length $\sim 1$
mm, $\Omega/2\pi$ is $\sim 10$ GHz. The modulation strength
$J/2\pi$ can be tuned at the order of $0.1-1$ GHz to make the
system operate in the weak modulation regime. Photonic crystal
fiber filled with a high-density atomic gas has the potential to
create a nonlinearity with $g/2\pi$ up to $\sim 1$ GHz in the
few-photon level \cite{venkataraman12}. Strong group velocity
dispersion is desired to bring the phase-mismatching condition
$|\Delta k L_A| = n\pi$, which can be simplified to $|\beta_2 L_A
\Omega^2| \sim 2\pi$. The necessary high dispersion $\beta_2 \sim
10^{-19}$ s$^2/$m is possible with  dispersion engineering in the
photonic crystal fiber \cite{zhu14} or by exploiting mode crossing
in coupled waveguides \cite{mia19}. The theoretical proposal here
may also be implementable in the microwave frequency range using
superconducting quantum circuits \cite{lee19}.

In summary, we show that a ring resonator undergoing dynamic
modulation can be used to achieve an interacting Hamiltonian where
the interaction is local along the synthetic frequency dimension,
provided that the ring incorporates $\chi^{(3)}$ nonlinearity, and
with a specific design of the group velocity dispersion of the
ring. Our work complements other emerging efforts in creating
locally interacting Hamiltonian in synthetic space, by showing
that such locally interacting Hamiltonian can be achieved in a
system that has been widely used in photonics. The possibility of
creating locally interacting Hamiltonian in the synthetic
frequency dimension is fundamentally important for future studies
of photon-photon interactions and many-body physics with synthetic
dimensions in photonics, with promising potentials for a variety
of quantum optical applications such as quantum information
science and quantum communication technology
\cite{bloch08,tpreview}.

We notice Ref. \cite{barbiero19} when we were in the final stage
of preparing this paper.

\begin{acknowledgments}
This paper is supported by National Natural Science Foundation of
China (11974245), National Key R\&D Program of China
(2018YFA0306301 and 2017YFA0303701), Natural Science Foundation of
Shanghai (19ZR1475700). This work is also partially supported by
the U.S. Air Force Office of Scientific Research Grant No.
FA9550-18-1-0379, a Vannevar Bush Faculty Fellowship from the U.
S. Department of Defense (Grant No. N00014-17-1-3030), as well as
the U.S. National Science Foundation Grant No. CBET-1641069.
\end{acknowledgments}

\end{document}